\input{epsf}
\documentstyle[aas2pp4,flushrt]{article}

\lefthead{Sarma, et al.}
\righthead{VLA OH and \ion{H}{1} Zeeman Observations of the NGC 6334 
complex}

\begin{document}

\title{VLA OH and \ion{H}{1} ZEEMAN OBSERVATIONS \\ OF THE 
NGC 6334 COMPLEX}

\author{A. P. Sarma\altaffilmark{1}, T. H. Troland\altaffilmark{1}, 
D. A. Roberts\altaffilmark{2}, and R. M. Crutcher\altaffilmark{2}}

\altaffiltext{1}{Physics \& Astronomy Department, University of Kentucky, 
Lexington KY 40506 \\ sarma@pa.uky.edu, troland@pa.uky.edu}

\altaffiltext{2}{Astronomy Department, University of Illinois, Urbana IL 
61801 \\ dougr@ncsa.uiuc.edu, crutcher@astro.uiuc.edu}

\begin{center}
{\em{Received 1999 September 24; accepted 1999 
November 24}}
\end{center}

\begin{abstract}

We present OH and \ion{H}{1} Zeeman observations of the NGC 6334 
complex taken with the Very Large Array. The OH absorption profiles 
associated with the complex are relatively narrow 
($\Delta\mbox{v}_{\rm FWHM}$ $\sim$ 3 km $\mbox{s}^{-1}$) 
and single-peaked over most of the sources. The \ion{H}{1} absorption profiles
contain several blended velocity components. One of the compact continuum 
sources in the complex (source A) has a bipolar morphology. The OH absorption 
profiles toward this source display a gradient in velocity from the northern 
continuum lobe to the southern continuum lobe; this velocity gradient likely 
indicates a bipolar outflow of molecular gas from the central regions to the 
northern and southern lobes. Magnetic fields of the order  of 200 $\mu$G have 
been detected toward  three discrete continuum sources in  the complex. Virial 
estimates suggest that the detected magnetic fields  in these  sources are of the 
same order as the critical magnetic fields required to support  the molecular 
clouds associated with the sources against gravitational collapse.
\end{abstract}


\section{INTRODUCTION}

\subsection{The Zeeman Effect and Magnetic Fields}

The importance of magnetic fields in the star formation process is now widely 
acknowledged (e.g., \cite{mou87}; \cite{sal87}; \cite{hgmz93}; \cite{mzgh93}; 
\cite{rmc99}). The Zeeman effect in radio-frequency spectral lines is the only 
viable method for measuring the strength of the  magnetic field in interstellar 
molecular clouds and star-forming regions. If the Zeeman splitting is smaller 
than the linewidth, only the line-of-sight component of  the magnetic field 
can be measured; this is always the case for  non-maser lines. 

In this paper we discuss high spatial resolution Zeeman observations in OH 
and \ion{H}{1} absorption toward the NGC 6334 complex.  In the remainder 
of \S 1, we review some details of the complex;  \S 2 contains details of the 
observations and reduction of the data.  In \S 3, we present the results of our 
observations, and \S 4 contains discussions about these results.

\subsection{The NGC 6334 Complex}

NGC 6334 is a giant molecular cloud complex and star-forming region at a 
distance of 1.7 kpc (\cite{nec78}). It lies  about 0$\arcdeg$.5 above the galactic 
plane and extends $\sim$30$\arcmin$ parallel to it. The complex has been the 
subject of extensive studies at a variety of wavelengths. \cite{rcm82} mapped 
the region at 6 cm wavelength with the VLA. They found six discrete 
continuum sources that lie along a ridge of radio emission parallel to the 
Galactic plane. Three of these sources had been known from previous single 
dish observations by \cite{sm69}. RCM82 named these sources A to F, in 
order of increasing right ascension (Figures 1 \& 2, this paper). \cite{mfs79} 
mapped the region with a 40$-$250 $\micron$ photometer and found six far 
infrared (FIR) sources. Of these, five lie along the ridge of radio emission; an 
additional weaker source lies  further south. The FIR sources were labeled I 
through VI. Another continuum source, I(N), was seen at millimeter and 
submillimeter wavelengths (\cite{cfg78} 1978; \cite{g82} 1982); I(N) lies to 
the north of FIR I and was not detected in the observations of \cite{mfs79}. 
\cite{ddw77} detected four CO peaks within an extended region of emission. 
The nomenclature in this complex has become confused. In this paper, we 
adopt the convention of RCM82 for radio continuum sources.

Among the many phenomena associated with star formation in this complex 
are water masers (\cite{mr80}), OH masers (\cite{gm87}), methanol masers 
(\cite{mb89}), bipolar outflows in ionized and molecular gas (\cite{drdg95} 
1995; \cite{bc90}; \cite{pm91}), and shocked $\mbox{H}_{2}$  emission 
(\cite{sh89a}).  Several authors have discussed a picture of sequential star 
formation in this complex (\cite{cfg78} 1978; \cite{mr80}) in which the 
central parts of the complex are more evolved than the edges.  This conclusion 
was mainly based on the distribution of indicators of star formation such as 
OH and $\mbox{H}_{2}\mbox{O}$ masers and CO, far-IR and 1 mm peaks.  
However, RCM82 point out that the picture is definitely more complex than 
simple sequential star formation.  For example, source E, which lies at the 
edge of the complex, is an evolved source.

\begin{deluxetable}{lllllcrrrrr}
\scriptsize
\tablenum{1}
\tablewidth{0pt}
\tablecaption{Observational Parameters for VLA Observations}   
\tablehead{
\colhead{Parameter}  &
\colhead{\ion{H}{1}}    &
\colhead{OH}}
\startdata
Date    &   1993 May 27    &   1996 Jan. 27   \nl
 &  1993 May 28    &  \nl   &   1993 Oct. 22   &    \nl   
Configuration   &   CnB, DnC     &    CnB    \nl
R.A. of field center (B1950)   &   $17^{h}$$17^{m}$$00^{s}$.0   &
    $17^{h}$$17^{m}$$00^{s}$.0    \nl
Decl. of field center (B1950)   &    $-35$\arcdeg$45$\arcmin$00$\arcsec$.0$
    &    $-35$\arcdeg$45$\arcmin$00$\arcsec$.0$    \nl
Total bandwidth (MHz)   &   0.78   &  0.19    \nl
No. of channels   &   128   &   128   \nl
Hanning smoothing   &   Yes   &   No   \nl
Channel Spacing (km $\mbox{s}^{-1}$)   &   1.28   &   0.27  \nl
Approx. time on source (hr)   &   12.6   &   4.3   \nl
Rest frequency (MHz)   &   1420.406   &   1665.402   \nl
   &    &   1667.359   \nl
FWHM of synthesized beam   &   35$\arcsec$ $\times$ 20$\arcsec$   &
   16$\arcsec$ $\times$ 12$\arcsec$   \nl
rms noise (mJy $\mbox{beam}^{-1}$)   &   &   \nl
\phm{abc}Line channels   &   10    &    8  \nl
\phm{abc}Continuum    &   9   &   7   \nl
\enddata
\tablecomments{1$\arcmin$=0.5 pc, assuming distance to NGC 6334 = 1.7 kpc}
\end{deluxetable}

The NGC 6334 complex has also been the subject of extensive recent 
observational studies. \cite{k97} and \cite{kth98} observed the complex in 
transitions of CO, CS, and  $\mbox{NH}_{\rm3}$ and in ionized carbon and 
neutral oxygen ([\ion{C}{2}] 158 $\micron$, [\ion{O}{1}] 145 $\micron$, and 
[\ion{O}{1}] 63 $\micron$). They found that the molecular emission in 
NGC 6334 shows a complex structure of filaments and bubbles, some of which 
are filled with photodissociated gas (\cite{kaas98}). \cite{bb95} observed the 
NGC 6334 complex in [\ion{C}{2}] 158 $\micron$ lines. They found that the 
\ion{C}{2} radiation is bright and widespread, with a general correlation 
between regions of intense \ion{C}{2} emission and warm dust and CO 
radiation.

\section{OBSERVATIONS AND DATA REDUCTION}

The observations were carried out with the Very Large Array (VLA) of the 
NRAO. \footnote{The National Radio Astronomy Observatory (NRAO) is 
operated by the Associated Universities, Inc., under cooperative agreement 
with the National Science Foundation.} The \ion{H}{1} observations were 
carried out in 1993 in the CnB and DnC configurations. The OH observations 
were carried out in January 1996 in the CnB configuration. Table 1 lists the 
important parameters of the \ion{H}{1} and OH observations.  For 
\ion{H}{1}, both circular polarizations  were observed simultaneously. For 
OH, both circular polarizations and both main lines (1665 and 1667 MHz) 
were observed simultaneously. In order to mitigate instrumental effects, a 
front-end transfer switch was used to reverse the sense of circular polarization
received at each telescope every 10 minutes.  Further, for the \ion{H}{1} 
observations, all calibration sources were observed at frequencies displaced 1 
MHz above and below the observing frequency for the source so that the 
Galactic \ion{H}{1} emission would not affect the calibration. Roberts, 
Crutcher, \& Troland (1995) describe very similar  observational techniques 
applied to S106.

The editing, calibration, Fourier transformation, deconvolution, and processing 
of the \ion{H}{1} and OH data were carried out using the Astronomical 
Image Processing System (AIPS) of the NRAO. Further processing of the data 
made use of the Multichannel Image Reconstruction, Image Analysis and 
Display (MIRIAD) system of the  Berkeley-Illinois-Maryland Array (BIMA).

OH masers in the field of view can have significant effects upon the Zeeman 
analysis because they are often highly circularly polarized.  Moreover, the 
effects of strong masers can be spread over the entire field of view in dirty 
maps owing to sidelobes of the synthesized beam.  Successful removal of 
maser effects is only feasible if the data are very well calibrated. Only in this 
case is the actual response in the maps to a point source (e.g., a maser) equal to 
the calculated point source response function  (the ``dirty'' beam).  The 
problem of maser contamination is especially acute in the NGC 6334 region 
where there are two strong masers in the field.  The strongest maser 
($\sim$400 Jy) coincides with source F (Fig. 2), the next strongest ($\sim$50 Jy) 
lies within the  shell-like region to the southwest of source A (\S 3.1; Fig. 2). 
There is also a weak  maser at 1665 MHz ($\sim$1 Jy) on the western edge of 
source A (Fig. 4a). Fluxes given are the sum of right and left circular 
polarizations at 1665 MHz. To remove maser effects  from the NGC 6334 OH 
data sets, we employed a several step process. First we performed a self 
calibration on the frequency  channel with the strongest maser radiation. This 
channel is at a velocity of  $-$8.8 km/s. Then we applied this self calibration 
solution to the entire  dataset. Next we generated an image  for each channel 
having maser radiation,  and we specified boxes around each maser in the 
image. Within each box, we  identified clean components that represented the 
maser radiation.  Next we  used the AIPS task UVSUB to  compute and remove 
from the uv dataset the  effects of these masers by using the clean components 
as the input model.  Finally, we re-generated the images for  each channel with  
the maser effects  removed and combined these images  with those of the other 
(non-maser affected) frequency channels.

\section{RESULTS}

\subsection{Continuum}

Figure 1 shows the 18 cm image of the discrete continuum sources in the 
NGC 6334 complex. Figure 2 shows the 21 cm grayscale continuum image 
of the entire complex, including the low-brightness shell-like sources in 
the extreme south that are not shown in Fig. 1. The 18 cm image has been 
made at higher  resolution (12$\arcsec$ $\times$ 9$\arcsec$) by uniformly
weighting the data. The compact radio continuum sources in this complex lie 
in a ridge that  extends from the northeast to the southwest; there is an 
underlying bed of continuum which extends $\sim$5$\arcmin$ to the 
northwest of the ridge.   The compact sources are labeled A$-$F, as described 
above (\S 1.2 and Fig. 1).  RCM82 and \cite{rcm88}  have discussed the nature 
and morphology of these  sources.  RCM82 concluded that all the compact 
sources, with the exception of  source B, are \ion{H}{2} regions.  
\cite{mgrb90} have shown that B is extragalactic. For the compact sources, 
our estimates of the integrated fluxes are in reasonable agreement (within 
about 10$\%$) with RCM82, except for source E, where we  obtain a flux of 
about 6 Jy, compared to their value of 12 Jy.

\begin{figure}[ptb]
   \centerline{\epsfxsize=2.7in\epsfysize=2.21in\epsfbox{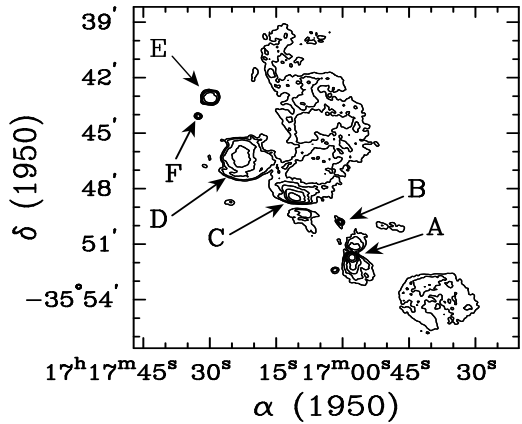}}
   \caption{{\small{Contour image of the continuum sources in the NGC 6334 
complex at 18 cm. Contour intervals are at 0.05, 0.1, 0.2, 0.4, 0.8, 1.6 Jy 
$\mbox{beam}^{-1}$. The beam is 12$\arcsec$ $\times$ 9$\arcsec$; the image 
has been made at a higher resolution by uniformly weighting the data.  The rms 
in the image is 7 mJy $\mbox{beam}^{-1}$. Note that the two weak shell-like 
structures in the extreme south of NGC 6334, shown in Fig. 2, are not included 
in this figure.}}}
\end{figure}

\begin{figure}[btf]
   \centerline{\epsfxsize=3.0in\epsfysize=2.21in\epsfbox{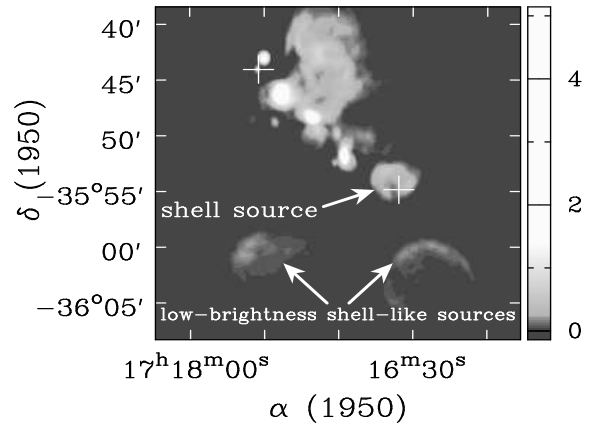}}
   \caption{{\normalsize{Grayscale radio continuum image of the entire NGC 6334 
complex at 21 cm. The beam is 35$\arcsec$ $\times$ 20$\arcsec$,  the rms in 
the image is 9 mJy $\mbox{beam}^{-1}$. The positions of the two strongest
OH masers are marked by the white crosses; the northeastern OH maser 
coincides with source F. The shell-like source about 4$\arcmin$ to the 
southwest of source A is designated as ``shell source'' in this 
figure.}} }
 \end{figure}

\begin{figure}[bt]
   \centerline{\epsfxsize=2.7in\epsfbox{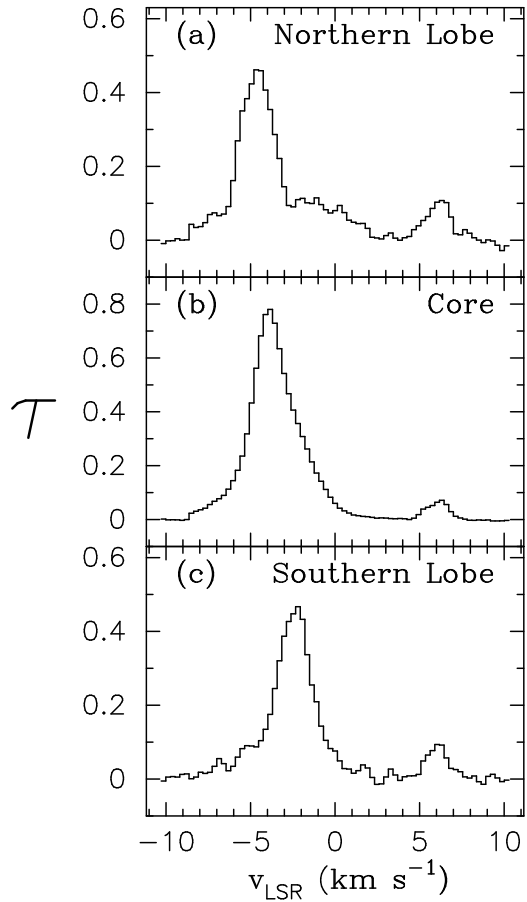}}
   \caption{{\scriptsize{OH 1667 MHz optical depth profile toward (a) the
northern lobe of source A at $\alpha$=$17^{h}$$16^{m}$$57^{s}$.8, 
$\delta$=$-$35$\arcdeg$51$\arcmin$09$\arcsec$; (b) the continuum peak of 
source A at $\alpha$=$17^{h}$$16^{m}$$57^{s}$.8, $\delta$=
$-$35$\arcdeg$51$\arcmin$42$\arcsec$; and (c) the southern lobe of source A 
at $\alpha$=$17^{h}$$16^{m}$$57^{s}$.5, $\delta$=
$-$35$\arcdeg$52$\arcmin$27$\arcsec$.}}}
\end{figure}

The bipolar morphology of source A (RCM 88) is clearly seen in the images
(Fig. 1 and 4a). In their higher resolution 6 cm image, RCM88 also see a 
discontinuous shell in the core of the source. The weak unresolved source 
G351.24$+$0.65 to the east of source A was first discovered by \cite{mgrb90}; 
\cite{drdg95} (1995) concluded that this is an optically thin \ion{H}{2}
region. There is also a shell-like source about 4$\arcmin$ to the southwest of 
A (Fig. 1, 2; designated as ``shell source'' in Fig. 2).  Parts of it coincide with 
the position of FIR source V.  \cite{jk99} conclude that the relative distribution 
of ionized, photodissociated, and molecular gas (as seen in radio continuum,
[\ion{C}{2}] 158 $\micron$, and CO 2$\rightarrow$1 emission respectively) 
toward this shell-like source conforms closely to an idealized model of  a
photodissociation region (PDR). In our 18 cm observations, the shell is almost
complete, except for an  opening in the south. An OH maser lies toward the 
south of this source, near  the break in the shell (Fig. 2).

Of the other main sources in the NGC 6334 complex, source C is extended and 
has a nonspherical appearance in the 6 cm image of RCM82. There is a steep 
decrease in the continuum intensity toward the south of this source (Fig. 1). In 
both our 18 cm and 21 cm images, there is a  low intensity source about 
1$\arcmin$ to the south of C.  Source D is extended, amorphous, and roughly 
spherical. Source E is  also extended and spherical. Source F is unresolved in 
our observations. RCM82 found it to have a ``nozzle-like'' morphology with 
a sharp decrease in emission to the west and a  smooth extension to the east. A 
very strong OH maser is coincident with the position of this source (\S 2; Fig. 2). 

Two weak shell-like structures appear to the south of the star-forming ridge 
in NGC 6334 (Fig. 2, where they are marked as ``low-brightness shell-like
sources''). These two sources are coincident with the bright southern parts of 
the optical nebula of NGC 6334 (\cite{kj99}). Also, part of  the western 
shell-like source is coincident with the extended FIR source VI of  
\cite{mfs79}. Since the faint radio shells are seen as bright objects in the red  
plate of the Palomar Sky Survey, the extinction toward these shells must be  
very low.

\subsection{OH \& \ion{H}{1} Absorption}

Optical depth profiles for the OH and \ion{H}{1} lines were calculated 
using the procedure described in \cite{rct95}. These OH and \ion{H}{1}
profiles all contain a  component at about $+$7 km $\mbox{s}^{-1}$ that 
likely arises in a foreground cloud. A cold \ion{H}{1} cloud was observed at 
this velocity toward the Galactic center by \cite{rc72} and \cite{cl84} and 
estimated to lie within 150 pc of the Sun. In all discussion that follows, we 
omit consideration of this foreground component and concentrate on negative 
velocity absorption that is associated with the NGC 6334 complex. The 
\ion{H}{1} profiles generally consist of numerous blended velocity 
components extending to velocities as negative as 
$-$25 km $\mbox{s}^{\rm -1}$ (e.g., Fig. 6). The only exception to this rule 
is toward source C. In this case the \ion{H}{1} profiles have a single relatively 
narrow component ($\Delta\mbox{v}_{\rm FWHM}$ $\sim$6 km 
$\mbox{s}^{\rm -1}$) centered at $-$6.6 km $\mbox{s}^{\rm -1}$.

The OH optical depth profiles have narrower less blended components with
center velocities in the range $-$2 to $-$7 km $\mbox{s}^{-1}$. Their
center velocities and linewidths are summarized in Table 2. We consider these
components to be associated with the NGC 6334 complex based on CO
and other molecular emission lines (DDW77; \cite{drdg95} 1995). Figure  3 
shows the profiles in OH 1667 MHz toward the core and lobes of source A; 
the profiles toward this source exhibit a north-south velocity gradient of 
about 3.4 km $\mbox{s}^{\rm -1}$ $\mbox{pc}^{\rm -1}$.  The 
\ion{H}{1} absorption toward source A is saturated; velocity structure is 
more difficult to quantify. However, the \ion{H}{1} profiles show evidence
of a  north-south velocity  gradient similar to that in OH. We do not detect 
any OH absorption  toward source C; however, detectable OH absorption is 
seen toward the low intensity continuum source about 1$\arcmin$ to the south 
of C. Considerable velocity structure is seen in OH absorption  profiles over 
source D west of the continuum peak; we do not detect any OH absorption to 
the east of the continuum peak.  Toward source E, the OH  absorption profiles 
are single-peaked and broad toward the northeast of the source, and  show two 
peaks toward the southwest.

\begin{figure}[tbp]
   \centerline{\epsfxsize=2.5in\epsfbox{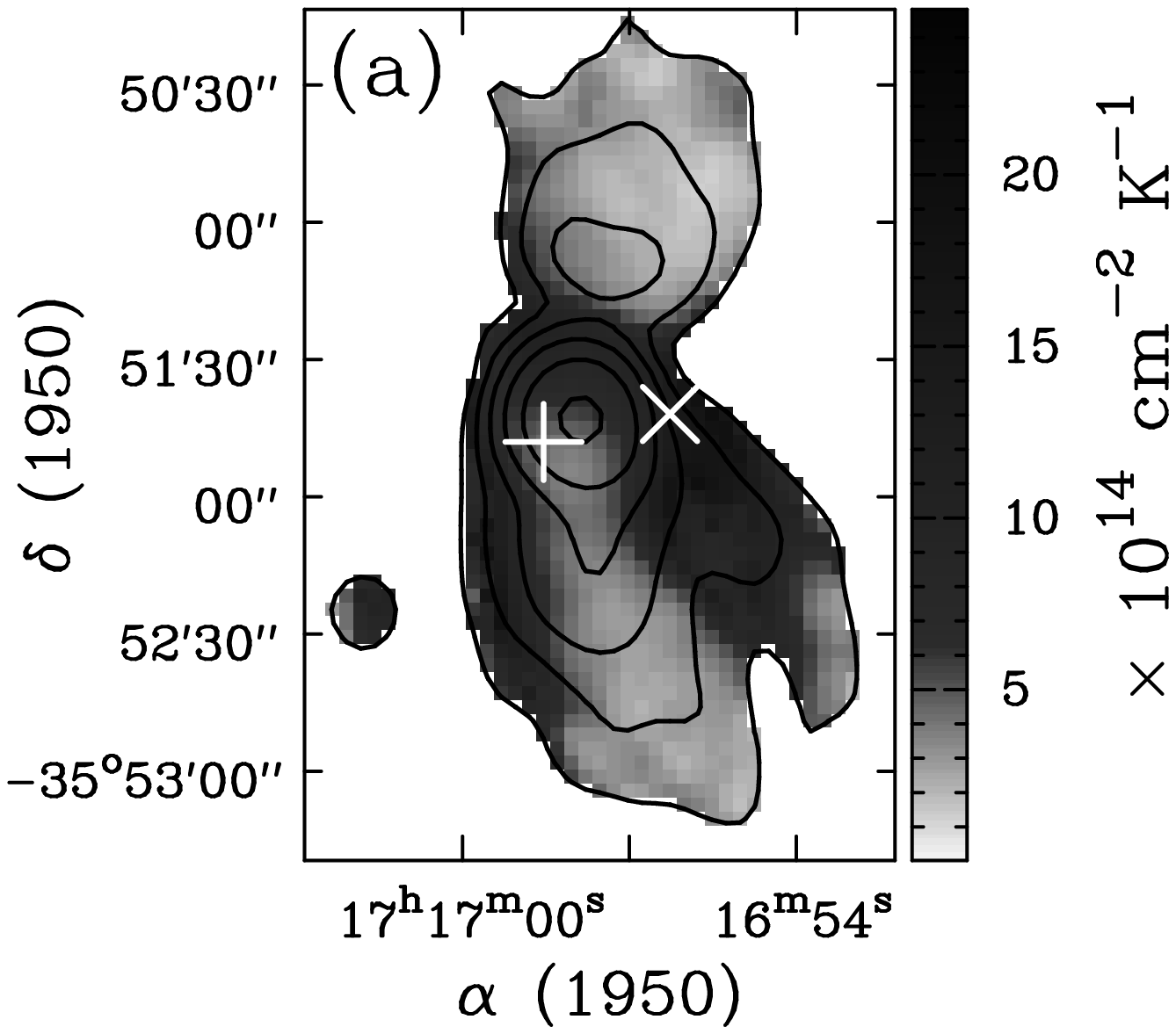}}
   \centerline{\epsfxsize=2.5in\epsfbox{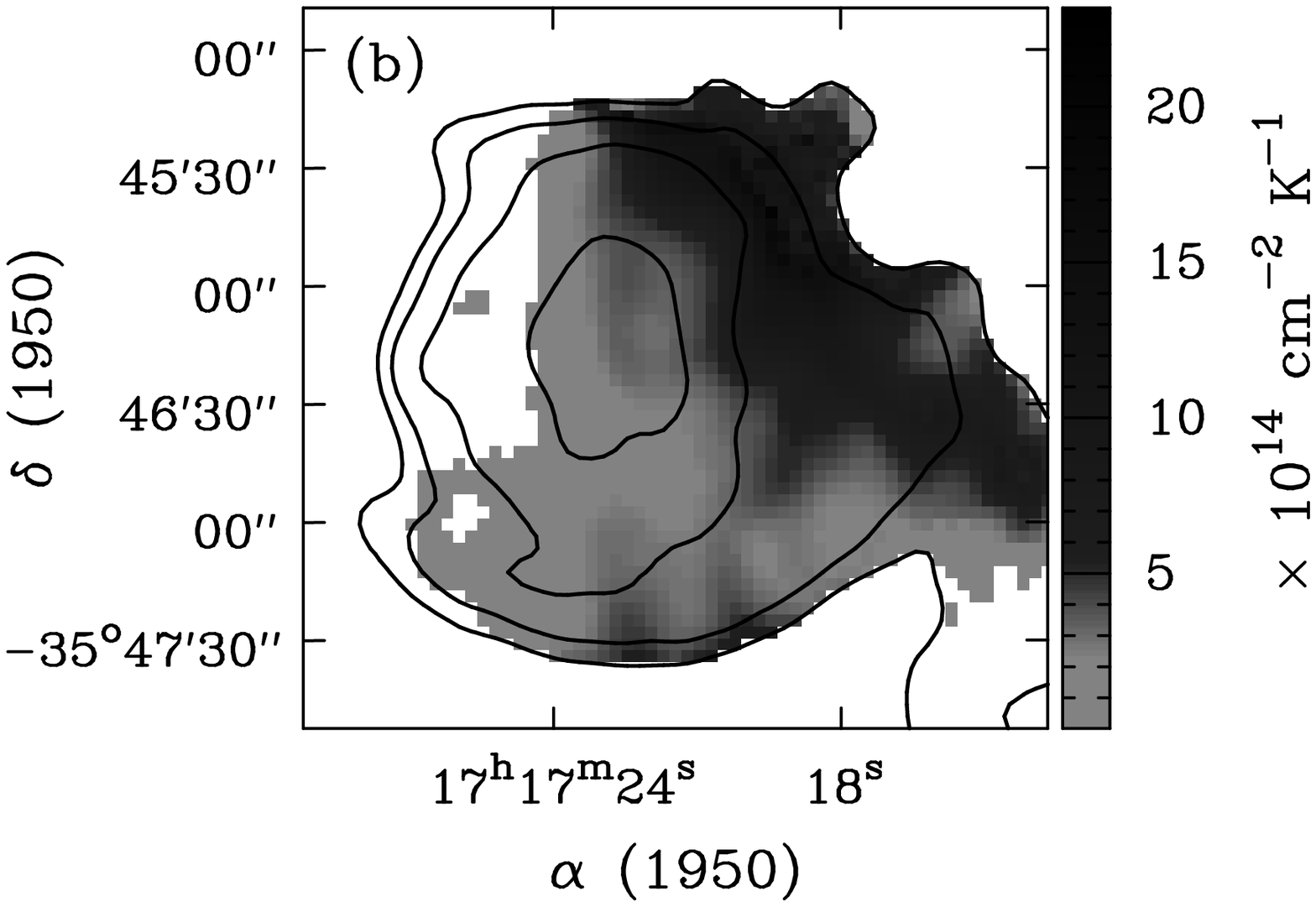}}
   \centerline{\epsfxsize=2.5in\epsfbox{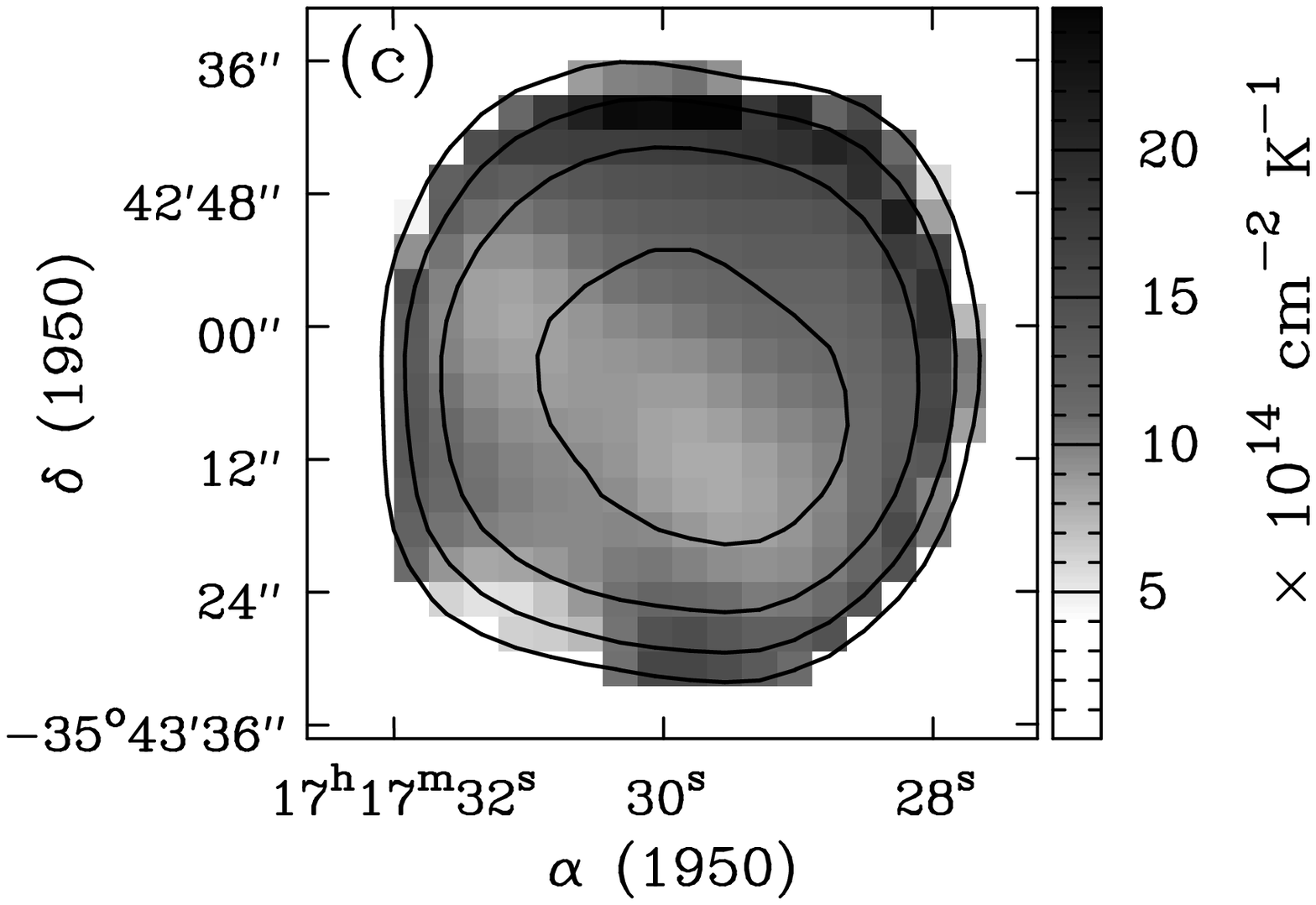}}
   \caption{{\scriptsize{(a) Grayscale plot of 
N(OH)/$\mbox{T}_{\rm ex}$ toward source A in the NGC 6334 complex. The 
contours represent the naturally weighted 18 cm continuum, contours are at 0.1, 
0.2, 0.4, 0.8, 1.6, 3.2 Jy $\mbox{beam}^{-1}$. The beam is 16$\arcsec$ 
$\times$ 12$\arcsec$. The `$+$' represents the position at which we  have 
measured the B-field. The `$\times$' represents the position of the $\sim$1 Jy 
1665 MHz maser on the western edge of source A. (b) Similar plot to (a), but 
toward source D.  The contours are at 0.1, 0.2, 0.4, 0.8 Jy 
$\mbox{beam}^{-1}$. (c) Similar plot  to (a), toward source E. The contours 
are at 0.1, 0.2, 0.4, 0.8 Jy  $\mbox{beam}^{-1}$.}}}
\end{figure}

The OH column density was determined using the relation
\begin{equation}
\mbox{N}_{\rm OH} / \mbox{T}_{\rm ex} = 
\mbox{C} \int {{\tau}_{\rm v} \mbox{dv}} 
\phm{abc}\mbox{cm}^{\rm -2}\phm{b}\mbox{K}^{\rm -1} 
\end{equation}
where $\mbox{T}_{\rm ex}$ is the excitation temperature of OH and the 
constant C = $4.1063 \times 10^{\rm14}$ and $2.2785 \times 10^{\rm14}$ 
$\mbox{cm}^{-2}$ (km $\mbox{s}^{-1}$)$^{-1}$ for the OH 1665 and 
1667 MHz lines respectively (\cite{c77}). Figure 4 shows plots of 
N(OH)/$\mbox{T}_{\rm ex}$ toward sources A, D, and E in OH at 1667 
MHz. In principle, variations in N(OH)/$\mbox{T}_{\rm ex}$ may arise
due to variations in the OH column density, or due to variations in 
$\mbox{T}_{\rm ex}$, or both. However, in our results and discussion, we
quote only variations in N(OH)/$\mbox{T}_{\rm ex}$, since the excitation
temperature cannot be measured based on absorption studies alone. Further
discussion about possible variations in $\mbox{T}_{\rm ex}$ appears in \S 4.2.
Toward A, there is an increase in N(OH)/$\mbox{T}_{\rm
ex}$ just  north of the continuum peak in the core; this ridge of enhanced OH 
spans the source from east to west. Also, we observe an increase in 
N(OH)/$\mbox{T}_{\rm ex}$ along the eastern and western boundaries of the 
northern and southern lobes of source A. This effect is especially striking in 
the southern lobe. Toward source D, there is an increase in 
N(OH)/$\mbox{T}_{\rm ex}$ going west from  the continuum peak (Fig. 4b) 
and no detectable OH absorption east of the continuum peak. 
N(OH)/$\mbox{T}_{\rm ex}$ is maximum in the northwestern part of this 
source.  Toward source E, N(OH)/$\mbox{T}_{\rm ex}$ appears to increase 
from the center to  the edge over most of the source (Fig. 4c).  However, 
nothing definite can  be said regarding the OH column density  at the 
southeastern edge, where the information has been blanked in many channels  
to remove remnant effects of the strong maser in source F.

Finally, we note the absorption profiles toward some of the other positions in 
the NGC 6334 complex. Absorption in both OH and \ion{H}{1} is seen 
toward the shell source to the southwest of A. OH  absorption is seen toward 
the highest intensity patches in the north and west of this shell. The OH lines 
toward these regions have a $\Delta\mbox{v}_{\rm FWHM}$ $\sim$ 3 km 
$\mbox{s}^{-1}$ and are centered near $\mbox{v}_{\rm LSR}$ $\sim$ 
$-$6 km $\mbox{s}^{-1}$. Broad \ion{H}{1} absorption profiles with 
several blended velocity components are also seen toward the two 
low-brightness shell-like sources in the extreme south of NGC 6334 (Fig. 2). 
Furthermore, OH and \ion{H}{1} absorption is also seen toward some of the 
brighter regions of continuum emission to the northwest of source D.

\begin{deluxetable}{ccccccccc}
\tiny
\tablenum{2}
\tablecolumns{6}
%%\tablewidth{0pc}
\tablecaption{Observed features of line profiles}   
\tablehead{
\colhead{}  &  
\multicolumn{2}{c}{Position}   &    \colhead{}   &
\multicolumn{2}{c}{OH (1667 MHz)\tablenotemark{(a)} }      \nl
\cline{2-3}    \cline{5-6}        \\
	\colhead{Source}   &   
	\colhead{R.A.}   &   \colhead{Decl.}   &    \colhead{}   &
	\colhead{$\mbox{v}_{\rm LSR}$}   &
	\colhead{$\Delta\mbox{v}_{\rm FWHM}$}     \nl
		\colhead{}   &
		\colhead{(B1950)}   &   \colhead{(B1950)}    &      
						\colhead{}   &
		\colhead{(km $\mbox{s}^{-1}$)}   &
		\colhead{(km $\mbox{s}^{-1}$)}    }   
\startdata
A   &      
$17^{h}$$16^{m}$$57^{s}$.8\phn     &
$-$35$\arcdeg$51$\arcmin$42$\arcsec$         &      &
$-$3.8   &   $\sim$3    &       \nl

C   &      
$17^{h}$$17^{m}$$10^{s}$.9\phn     &
$-$35$\arcdeg$48$\arcmin$30$\arcsec$        &      &
\nodata   &   \nodata    &       
 \nl

S. of C   &      
$17^{h}$$17^{m}$$08^{s}$.1\phn     &
$-$35$\arcdeg$49$\arcmin$24$\arcsec$        &      &
$-$2.5   &   $\sim$2    &       
%%&
\nl

D   &      
$17^{h}$$17^{m}$$23^{s}$.0\phn     &
$-$35$\arcdeg$46$\arcmin$15$\arcsec$        &      &
$-$5.8  &   $\sim$3  &       
%%&
\nl

D   &      
$17^{h}$$17^{m}$$18^{s}$.7\phn     &
$-$35$\arcdeg$46$\arcmin$39$\arcsec$         &      &
$-$4.7,  $-$1.4  &    $\sim$3,   $\sim$3    &       
%% &
\nl

D   &      
$17^{h}$$17^{m}$$17^{s}$.0\phn     &
$-$35$\arcdeg$46$\arcmin$51$\arcsec$        &      &
$-$3.0   &   $\sim$2    &       
%% &
\nl

E   &      
$17^{h}$$17^{m}$$30^{s}$.5\phn     &
$-$35$\arcdeg$42$\arcmin$57$\arcsec$        &      &
$-$5.5   &   $\sim$4    &       
%% &
\nl

E   &      
$17^{h}$$17^{m}$$29^{s}$.3\phn     &
$-$35$\arcdeg$43$\arcmin$15$\arcsec$        &      &
$-$6.9,  $-$2.2   &   $\sim$3,   $\sim$4 &       
%% &
\nl

\tablenotetext{(a)}{All OH profiles also show a component at positive
velocities which is not 
%%\\ \phm{abc}
shown in this table; see \S 3.2}
\tablecomments{\ion{H}{1} profiles show several blended velocity 
components between $-$25 
%%\\ \phm{abc}
and $+$14 km $\mbox{s}^{-1}$ 
toward all sources, except toward source C, where there 
%%\\ \phm{abc}
are two components; one at the positive velocity mentioned in (a) above
%% \\ \phm{abc}
and the other is at $\mbox{v}_{\rm LSR}$ = $-$6.6 km 
$\mbox{s}^{-1}$ and $\Delta\mbox{v}_{\rm FWHM}$ $\sim$6 km 
$\mbox{s}^{-1}$}

\enddata

\end{deluxetable}

\subsection{Magnetic Fields}

\begin{figure}[bp]
   \centerline{\epsfxsize=1.4in\epsfbox{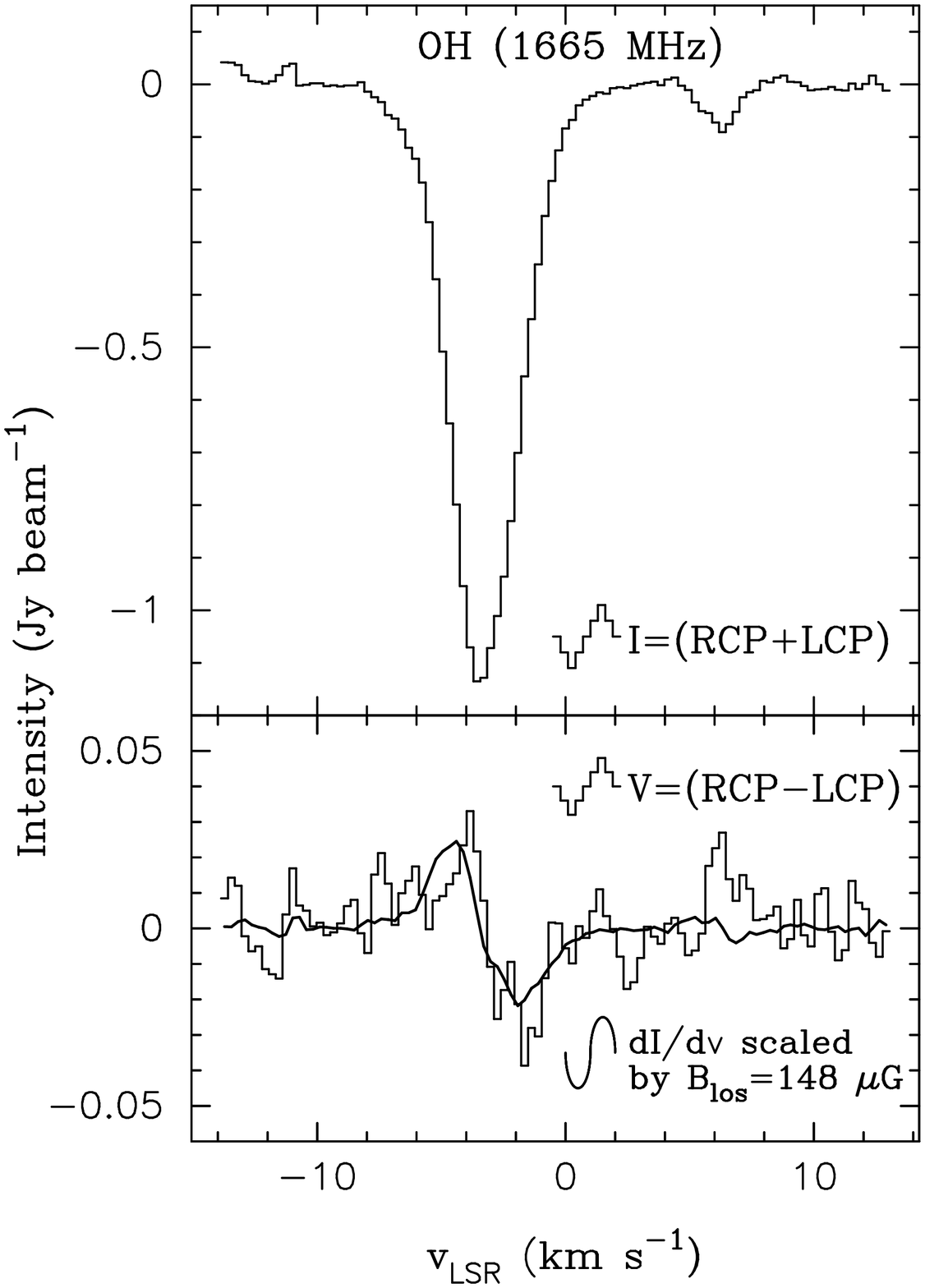}}
   \centerline{\epsfxsize=1.4in\epsfbox{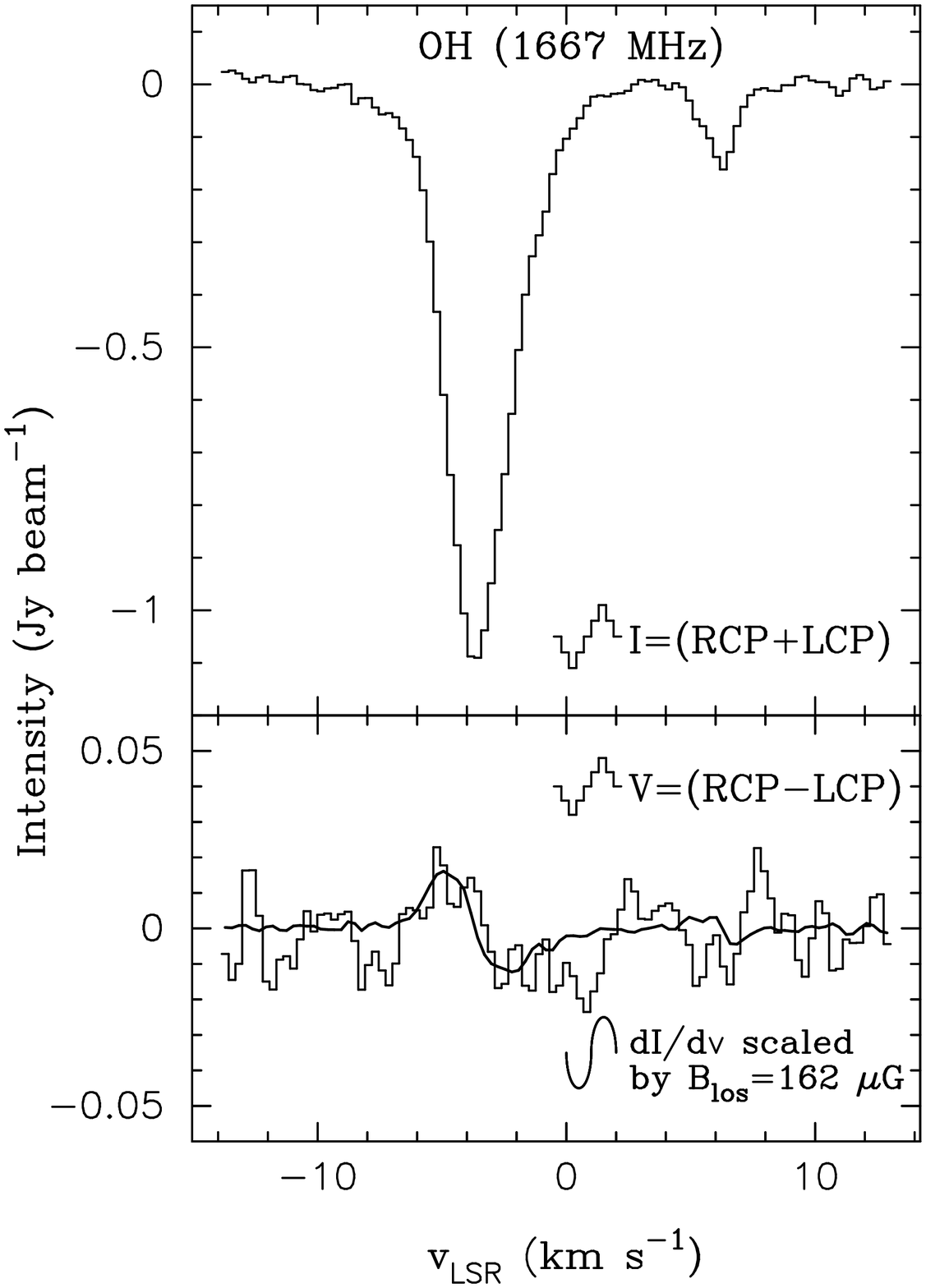}}
   \caption{{\scriptsize{(a) Stokes I (top-histogram) and V 
(bottom-histogram) profiles in the OH 1665 MHz line toward source A at the 
position marked by a `$+$' in Fig. 4a, $\alpha_{\rm 1950}$ = 
$17^{h}$$16^{m}$$58^{s}$.5, and $\delta_{\rm 1950}$ =
$-$35$\arcdeg$51$\arcmin$48$\arcsec$. The continuous line superposed on the 
V spectrum in the lower box is the derivative of I scaled by the derived value 
of $\mbox{B}_{\rm los}$ = 148 $\mu$G. (b) Same as (a), but for OH 1667 
MHz,  and a derived value of $\mbox{B}_{\rm los}$ = 162 $\mu$G.  }}}
\end{figure}

To determine magnetic field strengths using the Zeeman effect, we fitted a 
numerical derivative of the Stokes I spectrum to the Stokes V spectrum for
each pixel in the absorption line cube. The technique is described in 
\cite{rctg93}. The results of the fits, which give the line-of-sight component 
of the magnetic  field, $\mbox{B}_{\rm los}$, toward the various sources in 
the complex are listed in Table 3. We consider the results to be significant if 
the derived value of $\mbox{B}_{\rm los}$ is greater than  the 3$\sigma$ 
level. For OH, we have imposed a stronger condition $-$ the results are 
considered significant only if $\mbox{B}_{\rm los}$ is greater than the 
3$\sigma$ level for both 1665 and 1667 MHz lines. Field values that we 
believe to be significant are shown in bold face type in the table. In OH, 
significant detections of the magnetic field were made toward source  A. In 
\ion{H}{1}, significant detections of the magnetic field were made  toward 
source E and source D.  Figure 5 shows the  Stokes I and V profiles toward the 
core of source A (the position marked with a `$+$' in Fig. 4a) in  the 1665 and 
1667 MHz lines of OH, together with the  derivative of the I profile scaled by 
the fitted value of the magnetic field.  Figure 6 shows the Stokes I and V 
profiles and the scaled derivative of the I  profile in \ion{H}{1} toward source 
E.  By convention, a positive value of   $\mbox{B}_{\rm los}$ indicates that 
the field is pointing away from the observer.

\begin{deluxetable}{crrrrrrrrr}
\footnotesize
\tablenum{3}
\tablecolumns{8}
%%\tablewidth{0pt}
\tablecaption{Magnetic Field Values}   
\tablehead{
\colhead{}  &
\multicolumn{2}{c}{Position}  &    \colhead{}  &
\multicolumn{2}{c}{OH}   &    \colhead{}  &
\colhead{\ion{H}{1}}   \nl
\cline{2-3}  \cline{5-6}   \cline{8-8}   \nl   
	\colhead{Source}  &
	\colhead{R.A.}    &     \colhead{Decl.}      &    \colhead{}  & 
	\colhead{1665 MHz}  &
	\colhead{1667 MHz}  &
	\colhead{}   &
	\colhead{1420 MHz}	}
\startdata
A   &     $17^{h}$$16^{m}$$58^{s}$.5    &
$-$35$\arcdeg$51$\arcmin$48$\arcsec$      &    
&   \bf{148$\pm$20}   &   \bf{162 $\pm$33}  &        &   47$\pm$15    \nl
D    &     $17^{h}$$17^{m}$$22^{s}$.2    &
$-$35$\arcdeg$46$\arcmin$21$\arcsec$      &    
&   $-$60$\pm$46   &   $-$69$\pm$58   &     &    \bf{$-$93$\pm$13}   \nl
E     &     $17^{h}$$17^{m}$$30^{s}$.1    &
$-$35$\arcdeg$43$\arcmin$09$\arcsec$      &    
&  $-$263$\pm$78\tablenotemark{a}   &  $-$340$\pm$78\tablenotemark{a}
			&       &   \bf{$-$169$\pm$33}    \nl
       	& &  &	&  &   &	&	 \bf{$-$180$\pm$29}\tablenotemark{b}	
							\nl
NW of D   &     $17^{h}$$17^{m}$$08^{s}$.6    &
$-$35$\arcdeg$42$\arcmin$29$\arcsec$      &    
&   \nodata  & \nodata   &  &  169$\pm$55   \nl

\enddata

\tablenotetext{a}{from OH data convolved with 35$\arcsec$ beam}
\tablenotetext{b}{from \ion{H}{1} data convolved with a 35$\arcsec$ circular 
			beam for comparison with OH results}
\tablecomments{values in boldface type are significant detections (see \S 3.3)}

\end{deluxetable}

\begin{figure}[bt]
   \centerline{\epsfxsize=1.5in\epsfbox{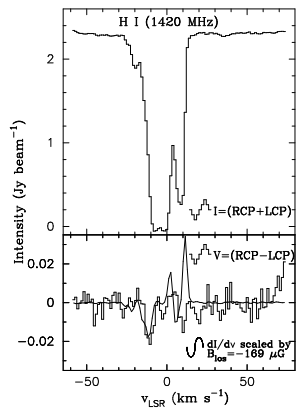}}
   \caption{{\scriptsize{Stokes I (top-histogram) and V (bottom-histogram) 
profiles in the \ion{H}{1} line toward source E at the position 
$\alpha_{\rm 1950}$ = $17^{h}$$17^{m}$$30^{s}$.1, and 
$\delta_{\rm 1950}$ = $-$35$\arcdeg$43$\arcmin$09$\arcsec$. The 
continuous line superposed on the V spectrum in the lower box is the derivative 
of I scaled by the derived value of $\mbox{B}_{\rm los}$ = $-$169 $\mu$G. }} }
\end{figure}

Toward source A, the OH data provide some tentative indications of field 
strength variations. Field strengths in the 1667 MHz line increase by a factor 
of two or more to the north of the continuum peak, in the direction of the 
east-west ridge of enhanced OH optical depth (Fig. 4a). However, weak maser 
contamination in this ridge at 1665 MHz prevented us from confirming this 
result in the 1665 MHz line. This maser coincides in velocity with one side of 
the absorption line, and its effect in the Stokes V profile cannot be removed
completely. We can, of course, exclude the channels contaminated by the 
maser from the fit, and compare the values of the magnetic field in 1665 and 
1667 MHz. However, higher resolution OH data will be needed to investigate 
such variations anyway since the  existing data provide only three to four 
independent resolution elements across  source A. Therefore, we defer 
examination of the spatial variation of the magnetic field in  source A to a 
later higher resolution study.

Toward source E, where we have a significant detection in \ion{H}{1}, we 
also find a possible detection in the 1665 and 1667 MHz absorption lines of 
OH. Our maser removal process has diminished the effect of the  masers to a 
large extent. However, it has proved impossible to completely remove the 
effect of the strong maser in source F in the Stokes V profile in a region so 
close to the maser. Therefore, in fitting for the magnetic field in source E, we 
have fitted over only those channels in the absorption line that are not affected 
by the maser. To improve our noise estimates, we have convolved our OH data 
with a  35$\arcsec$ beam. For comparison, we also convolved the  \ion{H}{1} 
data with a circular 35$\arcsec$ beam. Magnetic field values in the 1665 and 
1667 MHz OH data are consistent, and the \ion{H}{1} data reveal a 
significant field of similar order and the same sign. Finally, there is also a 
possible detection  in \ion{H}{1} toward an area of continuum to the 
northwest of source D. This fit is  barely at the 3$\sigma$ level, so we state it 
as an upper limit. If it is real, it represents a reversal in the sign of the field in
going from the region near D and E to the region northwest of D.

\section{DISCUSSION}

\subsection{Individual Sources in the NGC 6334 Complex}

\subsubsection{Source A}

The morphology of continuum emission from this source, including its bipolar
structure, has been described in \S 1.1. The peak of the FIR source IV lies near 
the peak of the radio source. Also associated with source A  are a 
Herbig-Haro-like object (\cite{ggd78}; \cite{boh92}), 
$\mbox{H}_{\rm2}\mbox{O}$ masers (\cite{rmgh80}; \cite{mr80}), and a  
CO peak (DDW77; \cite{pvg86}; K98).  RCM88 
suggested that the bipolar morphology of this region   results from the 
confinement of the \ion{H}{2} region by a flattened structure of gas and dust. 
This east-west confinement allows the ionized gas to escape  only to the north 
and south. Indeed, a 40$\arcsec$ resolution map of 100 $\micron$  optical 
depth  from \cite{hg83} shows an east-west elongation, which suggests a  
flattened structure of dust with an east-west extent of at least 
100$\arcsec$. Their map center coincides with the center of the ridge of 
enhanced N(OH)/$\mbox{T}_{\rm ex}$ in our data (\S 3.2 and Fig. 4a).
The east-west ridge of enhancement in our OH optical depth may 
be related to this disk. KJPB97 observed various transitions of CO and CS 
toward source A (HPBW of  CO J=1$\rightarrow$0 observations = 
53$\arcsec$). Their data show that source A is surrounded by a  
4$\arcmin$.4 $\times$ 1$\arcmin$.8 structure, indicating that the 
molecular  cloud is also flattened on much larger scales. 

The kinematical data for source A and its environs also offer insights into
the physical nature of this region. On the larger scale of 4$\arcmin$ (2 pc),
KJPB97 find an east-west velocity gradient of 2.4 
km $\mbox{s}^{\rm -1}$ $\mbox{pc}^{\rm -1}$ in the gas traced by
CO and CS. They conclude that a rotating disk of molecular gas surrounds
source A. Our OH absorption data also reveal an east-west velocity gradient 
of the same sign. Over a 35$\arcsec$ (0.3 pc) scale size, this gradient amounts
to 4 km $\mbox{s}^{\rm -1}$ $\mbox{pc}^{\rm -1}$. \cite{drdg95} (1995)
have observed ionized and neutral gas (H76$\alpha$ and H92$\alpha$
recombination lines, and $\mbox{H}_{\rm 2}$CO absorption) toward source 
A with a resolution of about 5$\arcsec$. They report some indications of 
east-west velocity gradients in the opposite sense of those cited above. 
KJPB97 have noted the disagreement between the sense of rotation as conveyed 
by their molecular data, and \cite{drdg95} (1995)'s recombination line and 
$\mbox{H}_{\rm 2}$CO absorption data.  However, our OH data agree in 
sign with KJPB97's data over a larger scale  size. The velocity gradients seen  
by \cite{drdg95} (1995)  over smaller size scales (about 15$\arcsec$) may,
therefore, reflect local velocity  inhomogeneities in the core of the source, 
rather than any bulk rotation  in the east-west direction.  

Evidence for north-south motions in the bipolar lobes comes from two sources.
\cite{drdg95} (1995) detect a north-south velocity gradient in H92$\alpha$
over a size scale of 100$\arcsec$, amounting to 19 
km  $\mbox{s}^{\rm -1}$ $\mbox{pc}^{\rm -1}$. They conclude that this 
indicates  a bipolar  outflow of ionized gas into the northern and southern 
lobes from the central source. They argue that this outflow lies nearly in the 
plane of the sky, based on the extended bipolar morphology and small 
velocity gradient. Our OH absorption data (\S 3.2  and Fig. 3) reveal a 
north-south gradient of 3.4 
km  $\mbox{s}^{\rm -1}$ $\mbox{pc}^{\rm -1}$ over a size scale of
80$\arcsec$. This suggests that the molecular gas traced by  OH absorption is 
also part of an outflow from the central regions to the northern and southern 
lobes. This north-south velocity gradient in OH has the same sign as the 
north-south gradient in H92$\alpha$ detected by \cite{drdg95} (1995). This 
correspondence in sign suggests that the OH gas may lie close to the 
neutral-ionized interface and be entrained by the outflow of the ionized gas.

The [\ion{C}{2}] 158 $\micron$ profile of BB95 (HPBW = 43$\arcsec$) 
toward FIR IV shows two peaks near $-$1 km $\mbox{s}^{-1}$ and $-$5 
km $\mbox{s}^{-1}$. BB95 note, however, that their spectra cannot be fitted 
by two Gaussian emission components. They mention a feature toward FIR IV 
with a velocity near $-$3 km $\mbox{s}^{-1}$ which is apparently due to 
self-absorption. We convolved our OH data with a 43$\arcsec$ beam for 
comparison with their data. Toward the position of FIR IV in source A, our 
OH profiles show an absorption feature near $-$3.7 km $\mbox{s}^{-1}$. 
Therefore, our data support the idea that the minimum between $-$1 and 
$-$5 km $\mbox{s}^{-1}$ in the [\ion{C}{2}] 158 $\micron$ profile is, 
indeed, self-absorption. The CO profiles of DDW77 and  PVG86 also show a 
peak near $-$3 km $\mbox{s}^{-1}$ toward FIR IV, and a  
$\Delta\mbox{v}_{\rm FWHM}$ $\sim$ 7 km $\mbox{s}^{-1}$. 

A significant $\mbox{B}_{\rm los}$ ($\sim$ 150 $\mu$G) was detected in 
OH 1665 and 1667 MHz absorption toward source A. We note that the 
magnetic field obtained from the formal fit in \ion{H}{1} is of the  same 
order although less than the magnetic field detected in OH. Due to the 
blending in velocity in the \ion{H}{1} gas, it is almost certain that optical 
depth effects will suppress any Zeeman signature in the V profile 
(\cite{sch86}),  which may explain why we do not have a more significant 
detection of $\mbox{B}_{\rm los}$ in \ion{H}{1} toward source A.

\subsubsection{Source D}

The extended, featureless source D (Fig. 4b) coincides with the FIR peak II of 
\cite{mfs79} and an $\mbox{H}_{\rm2}\mbox{O}$ maser (\cite{mr80}).
Based on their 6 cm radio continuum measurements, RCM82 found that a 
ZAMS O6.5 star would be  required to ionize this \ion{H}{2} region. 
\cite{shm89} found that IRS 24, which lies close to the center of the 
continuum source and has a  dereddened K magnitude corresponding to a 
ZAMS O6 star, is the dominant source  of excitation in this source. Source D 
is located at the edge of a hole  in the molecular gas emission, as seen in K98's 
CO data. Their models suggest  that the molecular hydrogen column density 
toward source D is the lowest of any source  in the NGC 6334 complex.  
\cite{sh89b} used near-infrared star counts to find the extinction through 
NGC 6334; their estimates also suggest that the extinction toward source D is 
less than that toward any other sources. This hole in the molecular  gas 
emission is clearly reflected in our high resolution image of source D (Fig. 4b),
where no OH absorption is detected east of the continuum peak. 

BB95 model their \ion{C}{2} profile toward FIR I (source F) as an emission 
component centered at $\mbox{v}_{\rm LSR}$ = $-$5.2 
km $\mbox{s}^{-1}$ and an absorption component near 
$\mbox{v}_{\rm LSR}$  = $-$1.6 km $\mbox{s}^{-1}$, and note that the 
source of the absorbing gas at this velocity is unclear. This absorption feature 
is also found near FIR II (toward source D). The CO profiles nearest to FIR II 
(of DDW77 with a 70$\arcsec$ beam and of PVG86 with a 1$\arcmin$.7 
beam) are broad ($\Delta\mbox{v}_{\rm FWHM}$ $\sim$ 7 km 
$\mbox{s}^{-1}$) but appear to be Gaussian and single-peaked near $-$3 
km $\mbox{s}^{-1}$. Our OH absorption profiles in this region show 
considerable  velocity structure (\S 3.2). However, when convolved with a 
43$\arcsec$ beam, our OH spectra show two components near 
$\mbox{v}_{\rm LSR}$ $\sim$ $-$5.0 and $-$1.6 km $\mbox{s}^{-1}$. 
This result suggests that both \ion{C}{2} velocity components identified by 
BB95 lie on the near side of the \ion{H}{2} region source D.

\subsubsection{Source E}

Source E is the northernmost compact radio continuum source in NGC 6334. 
RCM82 found that a ZAMS O7.5 star would be required to produce this 
\ion{H}{2} region. \cite{tpr96} detected at least 12 faint K-band sources 
toward source E; these sources were not detected in the J- or  H-bands. They 
suggested that these are a cluster of B0-B0.5 ZAMS stars which collectively 
ionize the \ion{H}{2} region.

The \ion{C}{2} profile of BB95 closest to source E is the one 1$\arcmin$ 
northwest of FIR I. The profile displays a peak near $\sim$ $-$8 km 
$\mbox{s}^{-1}$, and a lower level wing extends up to $\sim$ $+$2 km 
$\mbox{s}^{-1}$. Interestingly enough, the OH absorption profile toward the 
north of source E, when convolved with a 43$\arcsec$ beam for  comparison 
with BB95's data, is almost  identical to this \ion{C}{2} profile  in shape and 
extent, except that the peak of the absorption is at  $-$6.6  km 
$\mbox{s}^{-1}$. The \ion{H}{1} absorption is present at all these 
velocities. 

\subsection{Virial Estimates}

A principal goal of Zeeman effect measurements is to estimate the importance 
of the magnetic field to the dynamics and evolution of star forming regions 
like NGC 6334. Such estimates also require that other physical parameters of 
the regions be known, such as internal velocity dispersion, proton column 
density, radius, and total mass. C99 has summarized these concepts. Here, we
apply such a simple analysis of magnetic effects to NGC 6334 sources A, E, 
and D, especially source A for which physical quantities are best known. 

\begin{deluxetable}{cc}
\tablenum{4}
\tablecaption{Source A Parameters}   
\tablehead{
\colhead{Parameter}  &
\colhead{Value} }
\startdata
Radius (r)   &  0.8 pc   \nl
Mass (M)   &  2200 $\mbox{M}_{\rm \odot}$   \nl
$\mbox{N}_{\rm{H_{\rm2}}}$   &  4.8 $\times$ ${\rm 10}^{\rm22}$  
					$\mbox{cm}^{-2}$   \nl
$\mbox{n}_{\rm p}$   &   3 $\times$ ${\rm 10}^{\rm 4}$
					$\mbox{cm}^{-3}$   \nl
$\mbox{T}_{\rm k}$   &  40 K   \nl
$\mbox{$\Delta$v}$   &   2.6 km $\mbox{s}^{\rm -1}$    \nl
$\mbox{B}_{\rm los}$   &  150 $\mu$G  \nl

\enddata

\end{deluxetable}

First, we must estimate other relevant physical parameters. Column densities
can be estimated from molecular line data such as that reported here, subject
to several assumptions. For source A, we estimate column densities from our
own OH data. We use an average value of N(OH)/$\mbox{T}_{\rm ex}$ 
equal to 6 $\times$ $10^{14}$ $\mbox{cm}^{\rm -2}$ 
$\mbox{K}^{\rm -1}$ to calculate N(OH) toward this source. The excitation 
temperature cannot be measured based on absorption studies alone. We adopt 
$\mbox{T}_{\rm ex}$ = 40 K. K98 found $\mbox{T}_{\rm k}$ = 50 K 
toward source A based on CO studies. \cite{f87} found $\mbox{T}_{\rm k}$ 
$\sim$ 40 K from their ammonia studies. It is unlikely that 
$\mbox{T}_{\rm ex}$ remains constant over this source, and it may vary 
down to the dark cloud value of 10  K in the densest parts of the source. 
However, the results will scale with the temperature. The conversion ratio
$\mbox{N}_{\rm{H_{\rm2}}}$/N(OH)=2 $\times$ $10^{\rm6}$ was taken 
from \cite{rct95} for S106. This ratio is an order of magnitude higher than the 
ratio for dark clouds determined by \cite{c79}. However, we find 
N(OH)/$\mbox{A}_{\rm v}$ $\sim$ 5 $\times$ $\mbox{10}^{\rm 14}$, 
where we have used $\mbox{A}_{\rm v}$ = 50 mag., based on PVG86's 
estimate of $\mbox{A}_{\rm v}$  $\sim$ 40 mag. toward a position about 
1$\arcmin$ to the southeast of A. Using the  standard conversion  ratio of
$\mbox{N}_{\rm{H_{\rm2}}}$/$\mbox{A}_{\rm v}$  $\sim$ 
${\rm 10}^{\rm 21}$, we obtain $\mbox{N}_{\rm{H_{\rm2}}}$/N(OH) 
$\sim$ 2 $\times$ $10^{\rm6}$ in agreement with \cite{rct95}, which 
justifies our use of the \cite{rct95} value  rather than the dark cloud value. 
Using this conversion ratio, we obtain a value of  
$\mbox{N}_{\rm{H_{\rm2}}}$ = 4.8 $\times$ $\mbox{10}^{\rm 22}$ 
$\mbox{cm}^{\rm -2}$. Note that this value compares well with the value of 
$\mbox{N}_{\rm{H_{\rm2}}}$= 1.3  $\times$ $\mbox{10}^{\rm 23}$ 
$\mbox{cm}^{\rm -2}$ obtained by KJPB97 based on their CO excitation 
models, and with K98's value of 8.9 $\times$ $\mbox{10}^{\rm 22}$ 
$\mbox{cm}^{\rm -2}$ based on CO integrated  intensity. Next, the most 
problematic parameter is the radius of the  molecular cloud. K98 report a 
range of values for the radius based on their CO and  CS emission studies. 
Using the average value of $\mbox{N}_{\rm{H_{\rm2}}}$  = 4.8 $\times$ 
$\mbox{10}^{\rm 22}$ $\mbox{cm}^{\rm -2}$ and different values  for the 
radius, we estimated the mass and proton density  toward source A. The best 
value of the radius was then taken to be that for which  the mass of the 
molecular cloud most closely matched the mass obtained by KJPB97  from 
their CO data. Further, this adopted value of the radius (0.8 pc) matches well 
with the radius of the molecular cloud in $^{\rm 13}\mbox{CO}$   
2$\rightarrow$1 (0.8 pc) and CS  3$\rightarrow$2 (0.7 pc) from  KJPB97. 
Results of our analysis for source A are given in Table 4. Note that our value 
for the proton density is $\sim$ 2 higher than their value. For source E, where 
some of  the above parameters are not available, we adopt a value of 
$N_{\rm{H_{\rm2}}}$ =  6.3 $\times$ $10^{\rm22}$ 
$\mbox{cm}^{\rm -2}$ from K98, as derived from  CO excitation model 
calculations. Similarly, for source D, we use their value of  
$N_{\rm{H_{\rm2}}}$ = 2.2 $\times$ $10^{\rm22}$ 
$\mbox{cm}^{\rm -2}$.

The most straightforward estimate of the importance of the magnetic field 
comes from the relation :
\begin{equation}
\mbox{B}_{\rm S,crit} = 5 \times {10^{\rm -21} 
\phm{b}\mbox{N}_{\rm p}} \phm{abc}\mu \mbox{G}
\end{equation}
where  $\mbox{N}_{\rm p}$ is the average proton column density of the 
cloud, and $\mbox{B}_{\rm S,crit}$ is the average static magnetic field in 
the cloud that would  completely support it against self gravity. Note that the 
magnetic field in a cloud can be described in terms of a static component 
$\mbox{B}_{\rm S}$, and a  time-dependent component 
$\mbox{B}_{\rm W}$. It is likely that the  Zeeman effect primarily  samples 
the static component (see \cite{btrc99}, and references therein). If the actual 
field in the cloud is comparable to  $\mbox{B}_{\rm S,crit}$, then the field 
can be judged dynamically important to the region. For source A, we find 
$\mbox{B}_{\rm S,crit}$ = 500 $\mu$G, compared to an estimated total 
field strength of 300 $\mu$G. Following C99, we have used total magnetic 
field strength B equal to 2 times  $\mbox{B}_{\rm los}$, and 
$\mbox{$\vert$B$\vert$}^{\rm 2}$ equal to 3 times 
${\vert \mbox{B}_{\rm los}}\vert ^{\rm 2}$. For source E, we find 
$\mbox{B}_{\rm S,crit}$ = 600 $\mu$G, compared to an estimated total 
field strength of 400 $\mu$G. Again, for source D, we find 
$\mbox{B}_{\rm S,crit}$ = 220 $\mu$G, compared to an estimated total field 
strength of 190 $\mu$G. Therefore, subject to the obvious uncertainties in 
such estimates, we judge the magnetic field in each region to be comparable to 
if  slightly less than the critical field. In all three regions, the magnetic field 
should be dynamically significant, providing an important source of support 
against self gravity.

\begin{deluxetable}{cc}
\tablenum{5}
\tablecaption{Source A : Derived Values and Virial Estimates}   
\tablehead{
\colhead{Parameter}  &
\colhead{Value} }
\startdata
$\sigma$/c   &  2.9   \nl
$\sigma$/$\mbox{v}_{\rm A}$   &  0.3   \nl
$\mbox{$\beta$}_{\rm p}$   &  0.03   \nl
${[\mbox{M}/{{\Phi}_{\rm B}}]}_{obs/crit}$  &   1.6   \nl
$\cal{W}$    &  3.7 $\times$  ${\rm 10}^{\rm 47}$  ergs  \nl
$\cal{T}$/$\cal{W}$   &   0.22    \nl
$\cal{M}$/$\cal{W}$   &  0.27  \nl

\enddata
\tablecomments{$\sigma$ is the line velocity dispersion, c is the speed of sound,
$\mbox{v}_{\rm A}$ is the Alfv\'{e}n velocity, 
$\mbox{M}/{{\Phi}_{\rm B}}$ is the mass to magnetic flux ratio, 
$\cal{W}$ is the virial gravitational energy, $\cal{T}$ is the virial kinetic 
energy, and $\cal{M}$ is the virial magnetic energy; all terms are defined in 
C99.}

\end{deluxetable}

Other parameters of magnetic significance for source A can be inferred from 
the data of Table 4; they are shown in Table 5. All symbols used in the 
discussion that follows are described in Table 5. We find $\sigma$/c =2.9, 
indicating that the motions in the  cloud are  supersonic. This is a common 
result in molecular clouds; C99 has found that motions are supersonic by about 
a factor of 5. However, the ratio $\sigma$/$\mbox{v}_{\rm A}$ indicates 
that they are sub-Alfv\'{e}nic, suggesting that these motions may be the result of 
Alfv\'{e}n waves in the cloud.  The parameter  ${\rm \beta}_{\rm p}$ $\approx$ 
0.03, close to the average value of 0.04 determined by C99, suggesting that the
magnetic fields are important, and that the magnetic pressure dominates over 
the thermal pressure. The observed mass-to-magnetic flux ratio with respect 
to the critical value is 1.6, indicating that the cloud is magnetically 
supercritical. Also, $\cal{T}$ $\approx$ $\cal{M}$. Thus, the kinetic and 
magnetic energy densities are in approximate equilibrium, which would be 
expected if the magnitude of the fluctuating part of the magnetic field was  
approximately equal to the magnitude of the static field (C99). Finally, we 
find that 2$\cal{T}$ $+$ $\cal{M}$ $\approx$ 0.7$\cal{W}$. Hence, since 
the external pressure  term, which is not included here, will act in the same 
sense as the gravitational term, the cloud is in approximate virial equilibrium.

\section{CONCLUSIONS}

1). We have observed OH and \ion{H}{1} in absorption toward NGC 6334 
with the VLA. The OH absorption profiles are relatively narrow 
($\Delta\mbox{v}_{\rm FWHM}$ $\sim$ 3 km $\mbox{s}^{-1}$) 
and single-peaked toward most of the sources. Toward source A, the OH 
profiles display an east-west velocity gradient; this gradient has the same sign 
as the velocity gradient detected in molecular emission lines by KJPB97. The 
OH profiles toward source A also display a north-south velocity gradient 
which is in agreement with a similar gradient discovered in H92$\alpha$ 
recombination lines, and which likely indicates a  bipolar outflow of 
material into the lobes seen in the continuum source. No OH absorption is 
detected toward the continuum peak of source C itself. Toward source D, the 
OH profiles show considerable velocity structure west of the continuum peak; 
no OH absorption is detected east of the continuum peak. The \ion{H}{1} 
profiles toward NGC 6334 contain several blended velocity components, 
except toward source C, where the \ion{H}{1} profile at negative velocities is 
almost single-peaked.

2). We have calculated N(OH)/$\mbox{T}_{\rm ex}$ toward the continuum 
sources. Toward source A, we observe an increase in 
N(OH)/$\mbox{T}_{\rm ex}$ in an east-west ridge north of the continuum 
peak; there is also an increase toward the eastern and western boundaries of 
the northern and southern lobes. 

3). Magnetic fields of the order of 200 $\mu$G have been detected toward 
some of the sources in this complex. In OH, we have significant detections of 
magnetic fields toward source A, and possibly a significant detection toward 
source E. In \ion{H}{1}, we have significant detections toward sources E and 
D. There may also be a detection toward an area of continuum to the 
northwest of source D. Also, $\mbox{B}_{\rm los}$ obtained from the 
formal fit for source A in \ion{H}{1} is of the same order as 
$\mbox{B}_{\rm los}$ detected in OH absorption toward source A; 
similarly, $\mbox{B}_{\rm los}$ obtained from the formal fit for source D 
in OH matches $\mbox{B}_{\rm los}$ detected in \ion{H}{1} absorption 
toward this source.

4). We have three or four independent beams across source A; measurements at 
different positions in 1667 MHz suggest that the magnetic field increases 
going north of the continuum peak. The signs of the detected magnetic fields 
are opposite in going from sources E and D (where we have a detection in 
\ion{H}{1} absorption) to source A (where we have a detection in OH). If the 
field detected in \ion{H}{1} to the northwest of D is real, there is also a sign
reversal in the field in going from D and E to the position northwest of D.

5). We have used various observed and derived parameters to study the 
implications of the detected magnetic fields toward sources A, E, and D. In all 
cases, it appears that the detected fields are of the order of the critical field 
needed to support the molecular cloud associated with that source against 
gravitational collapse. We have also compared various derived parameters for 
source A with the published  values of those parameters determined from an 
ensemble of clouds by C99.

\acknowledgements

APS acknowledges a fellowship from the Center for Computational Sciences 
(CCS) at the University of Kentucky, and would like to thank John Connolly, 
Director, CCS, for providing much needed computational support. THT 
acknowledges NSF grant AST 94-19220; RMC acknowledges NSF grant 
AST 98-20641. This work has  made use of the NASA Astrophysics Data 
System (ADS) astronomy abstract service.

\end{document}